\pdfoutput=1

\documentclass[prl,superscriptaddress,amsmath,amssymb,twocolumn]{revtex4-1}

\usepackage{graphicx,bm,amsmath}
\usepackage[usenames,dvipsnames]{xcolor}
\usepackage[colorlinks,bookmarks=false,citecolor=NavyBlue,linkcolor=Red,urlcolor=blue]{hyperref}
\usepackage[bbgreekl]{mathbbol}
\usepackage{xcolor}
\usepackage[normalem]{ulem}
\usepackage{braket}
\usepackage{bm}
\usepackage{esint}

\def\doi{http://dx.doi.org/}

\def\Tr{\operatorname{Tr}}
\newcommand{\be}{\begin{equation}}
\newcommand{\ee}{\end{equation}}
\newcommand{\bea}{\begin{eqnarray}}
\newcommand{\eea}{\end{eqnarray}}

\def\nn{\nonumber\\}

\def\pp{\upsilon}
\def\nn{\mathfrak{n}}

\def\ph{p_{\rm h}}
\def\pp{p_{\rm p}}
\def\opp{q}

\def\dd{\bm{\varrho}}

\def\nn{\mathfrak{n}}

\newcommand{\titleinfo}{Generalised hydrodynamics with dephasing noise}

\begin{document}

\title{\titleinfo}

\author{Alvise Bastianello}
\affiliation{Institute for Theoretical Physics, University of Amsterdam, Science Park 904, 1098 XH Amsterdam, The Netherlands}

\author{Jacopo De Nardis}
\affiliation{Department of Physics and Astronomy, University of Ghent, Krijgslaan 281, 9000 Gent, Belgium}
    
\author{Andrea De Luca}
\affiliation{Laboratoire de Physique Th\'eorique et Mod\'elisation (CNRS UMR 8089),
Universit\'e de Cergy-Pontoise, F-95302 Cergy-Pontoise, France}    

\begin{abstract}
We consider the out-of-equilibrium dynamics of an interacting integrable system in the presence of an external dephasing noise. In the limit of large spatial correlation of the noise, we develop an exact description of the dynamics of the system based on a hydrodynamic formulation. This results in an additional term to the standard generalized hydrodynamics theory describing diffusive dynamics in the momentum space of the quasiparticles of the system, with a time- and momentum-dependent diffusion constant. Our analytical predictions are then benchmarked in the classical limit by comparison with a microscopic simulation of the non-linear Schr\"odinger equation, showing perfect agreement. In the quantum case, our predictions agree with state-of-the-art numerical simulations of the anisotropic Heisenberg spin in the accessible regime of times and with bosonization predictions in the limit of small dephasing times and temperatures.   
\end{abstract}

\maketitle

\paragraph{\it Introduction. ---}
Recent advances in controlling and manipulating quantum matter \cite{PhysRevLett.120.243201,Simon2011,Vijayan186} have spurred the development of novel methods to study the out-of-equilibrium dynamics of many-body systems. On the one hand, numerical methods based on tensor network algorithms had astonishing achievements, extending their range of applicability to longer times and to a larger class of systems and protocols~\cite{white1992,Schollwoeck2011,Karrasch2012,Haegeman,leviatan2017quantum,Calabrese_2016,SciPostPhys.2.1.002,2018PhRvB..97b4307K}. On the other hand, for one-dimensional exactly solvable models, a new set of analytical tools have been devised to access the long-time stationary state, correlation functions and entanglement production~\cite{PhysRevLett.106.227203,Eisert2015, alba2017entanglement,10.21468/SciPostPhys.4.6.037,2002.09527}.
For homogeneous integrable systems evolving under a time-independent Hamiltonian, it is now completely understood how to express the long-time evolution once the local and quasilocal conserved charges of the model have been classified~\cite{PhysRevLett.115.157201, Ilievski_2016}: 
their expectation values uniquely determine the generalised Gibbs ensemble (GGE) \cite{rigol2007relaxation} describing the late-times stationary state.

More recently, generalized hydrodynamics (GHD) provided an efficient framework to study integrable systems prepared in inhomogeneous states~\cite{PhysRevX.6.041065, Bertini16}. It progressed at a fast pace leading to several extensions~\cite{doyonfleagas,PhysRevB.96.115124,PhysRevB.96.081118}, analytic results~\cite{collura2018analytic}, applications~\cite{10.21468/SciPostPhys.7.1.005,Bertini2018, Ilievski17,10.21468/SciPostPhys.6.6.070,SciPostPhys.3.6.039,10.21468/SciPostPhys.5.5.054,10.21468/SciPostPhys.8.1.007}, studies of classical systems~\cite{10.21468/SciPostPhys.4.6.045, Doyon_2017,doi:10.1063/1.5096892,Bulchandani_2019} and even experimental confirmations~\cite{PhysRevLett.122.090601}. Further developments have included diffusive corrections~\cite{PhysRevLett.121.160603, 10.21468/SciPostPhys.6.4.049,PhysRevLett.122.127202,PhysRevB.98.220303,PhysRevLett.120.164101,1911.01995}, predictions beyond integrability~\cite{friedman2019diffusive, Bastianello_2019} and quantum fluctuations~\cite{ruggiero2019quantum}, and have extended its applicability to additional protocols, including space-time dependent forces \cite{doyon2017note} and interactions~\cite{PhysRevLett.123.130602}. 

However, a crucial aspect when comparing to real-world experiments is that interaction with the external environment will eventually affect the unitary evolution of the system. Modelling the open dynamics of a quantum system is a notoriously difficult problem~\cite{breuer2002theory} as there is not a unique way to incorporate the external degrees of freedom while first-principle constructions often lead to hardly treatable formulations~\cite{PhysRevLett.113.150403}. An important simplification occurs for setups where the correlation time of the bath can be neglected compared with the scale of the system itself. In this case, one can assume Markovianity and consistency with the laws of quantum mechanics restricts the possible form of open evolution to the so-called Lindblad equation.
In practice finding exact solutions for its dynamics is a difficult task, and it constitutes an active subject of research. Notable progresses were done in quadratic Fermi systems~\cite{Prosen_2008}, integrable Linbladians \cite{medvedyeva2016exact, rowlands2018noisy, PhysRevB.99.224432, aleks2019yangbaxter,PhysRevB.99.174303}, and by means of mappings to classical stochastic systems \cite{Bernard2019,2001.04278,1912.08458}.
At the leading order, the effect of the environment is to induce phase fluctuations between different portions of the system, 
without locally exchanging energy or other conserved quantities (although global heating is possible due to the interactions between different regions).
This \textit{dephasing} is described by a Lindbladian whose jump operators are Hermitian.    In this Letter, we introduce a general framework where the dynamics of an integrable model subject to the dephasing noise can be studied exactly. 
In particular, we consider the case where a fluctuating environment is locally coupled to any local operator, focussing primarily on local conserved charges.
In the spirit of GHD, we derive, in the long-wavelength limit, a compact evolution equation for the local stationary state, which admits a simple interpretation in terms of diffusion in the momentum space of the relevant quasiparticles. 

\paragraph{\it Noise and dephasing model.
---}
We consider generic homogeneous Bethe-Ansatz integrable Hamiltonians $\bm{\hat H_0}$. For definiteness, we focus on discrete systems whose sites are indexed by $j = 1,\ldots, L$, with a finite local Hilbert space (e.g. spin chains), although the discussion can be extended to other settings (see below). We assume that the evolution is described by the non-integrable Hamiltonian
\begin{equation}\label{eq:defModel}
 \bm{H}_{\eta} = \bm{H}_0 + \sum_j \eta_j(t) \bm{O}_j,
\end{equation}
where the second term encodes the dephasing noise, with $\langle \eta_j(t) \eta_{j'}(t') \rangle = \gamma F(j-j') \delta(t-t')$. The parameter $\gamma$ controls the intensity of the noise, while the function $F(x)$ its spatial correlation. 
The operator $\bm{O}_j$ is assumed to have (quasi)local support around the site $j$. 
The quantum dynamics of the model is then described as the solution of the Schr\"odinger equation
\begin{equation}
\label{stochschr}
\frac{ d\ket{\psi}}{dt} = -\imath \bm{H}_{\eta} \ket{\psi},
\end{equation}
which is a stochastic differential equation (SDE). Note that it involves a multiplicative noise term and the Stratonovich convention is assumed here~\cite{gardiner1985handbook} (see also the Supplementary Material (SM) \footnote{Supplementary material at [url] for details about i) convention with Stochastic equations; ii) Derivation of the GHD description of the dephasing model}). 
The noise-averaged density matrix $\dd = \overline{\ket{\psi}\bra{\psi}}$ satisfies the Lindblad equation
\begin{equation}
\label{eq:lindblad}
 \dot \dd = 
 -\imath [\bm{H}_0, \dd] - \frac \gamma 2  \sum_{j,j'} F(j-j') [\bm{O}_j, [\bm{O}_{j'}, \dd]].
\end{equation}
In general, solving Eq. \eqref{eq:lindblad} for a many-body system is even harder than its pure dynamics. For short-range noise, i.e. $F(j-j') \to \delta_{j,j'}$, a few solvable cases have recently been discovered: when $\bm{H}_0$ describes non-interacting spinless fermions and $\bm{ O}_j$  denotes their on-site occupation number, Eq.~\eqref{eq:lindblad} was shown to be related to the integrable Fermi-Hubbard model~\cite{medvedyeva2016exact}; other integrable 
examples have been classified in Ref. \cite{aleks2019yangbaxter}. Moreover the case $\bm{O} = S^z$ in the XXZ spin chain was recently studied in \cite{SciPostPhys.3.5.033,10.21468/SciPostPhys.6.4.045} in the limit $\gamma \to \infty$ and for $\delta-$ correlated noise. Here, instead, we focus on the opposite limit where the correlation $F(j-j')$ is flat within the correlation length $\ell$ and smoothly decays for  $|j-j'|\gg \ell$.

\paragraph{Hydrodynamics description. ---}
 Let us briefly describe the dynamics for $\gamma = 0$. Since $\bm H_0$ is integrable, there exists an infinite set of conserved quantities $\bm{Q}^{(\alpha)}$, $\alpha = 1,\ldots$ commuting with the Hamiltonian $[\bm{Q}^{(\alpha)}, \bm{H}_0] = 0$. Starting from an initial density matrix $\bm{\varrho}_0$, the unitary evolution preserves all the conserved quantities of the system $\bm{Q}^{(\alpha)}$
and induces equilibration to the GGE pinned down by such initial values~\cite{PhysRevLett.115.157201}. In practice, it is convenient to encode the GGE by introducing the root density of the quasiparticles $\rho(\lambda)$ \cite{takahashi2005thermodynamics}, defined such that $L \rho(\lambda) d\lambda$ equals the number of quasiparticles with rapidities $\in [\lambda, \lambda + d\lambda)$. Quasiparticles are conserved modes and their dynamics is fully encoded in the scattering shift $T(\lambda,\lambda')$ for any integrable system. For simplicity, here we consider a single quasiparticle species, the generalizations being straightforward. 
The rapidity $\lambda$ parametrises the state of each quasiparticle, such that, in the thermodynamic limit
\begin{equation}
\label{eq:chargerhodual}
\lim_{L \to \infty} \frac{ \Tr[\dd_0 \bm{Q}^{(\alpha)}]}{L} =  \int d\lambda \rho(\lambda)q^{(\alpha)}(\lambda)  \equiv \braket{\rho| \bm{Q}^{(\alpha)} | \rho},
\end{equation}
where the functions $q^{(\alpha)}(\lambda)$ are the single-particle eigenvalues associated to the $\alpha$--th charge. Eq.~\eqref{eq:chargerhodual} establishes the correspondence between a complete set of charges and the root density. In the last equality, we employed a generalized microcanonic ensemble to select a pure macrostate $\ket{\rho}$ representative of the root density $\rho(\lambda)$~\cite{PhysRevLett.110.257203}. 

Now, we turn on the weak dissipative term in Eq.~\eqref{eq:lindblad} and we assume that the system remains always in a GGE representative state $\ket{\rho(t)}$ which evolves in time. In order to get the evolution equation for the root density, we look at the time variation of the expectation values of the charges. We replace $\bm{\varrho} \to \ket{\rho(t)}\bra{\rho(t)}$ in the right-hand side of Eq. \eqref{eq:lindblad} and we obtain
\begin{equation}
\label{eq:chargevar}
\lim_{L \to \infty} \frac{\Tr[\bm{Q}^{(\alpha)} \dot{\bm{\dd}}]}{L} = 
  \gamma  \sum_{e} \Delta Q_e^{(\alpha)} \hat F(\Delta P_e) |\bra{\rho} \bm{O} \ket{\rho; e}|^2.
\end{equation}
Here, we assume the observable $\bm O$ is number conserving, hence we inserted a sum over the tower of all the possible \textit{particle-hole} excitations $\ket{\rho, e}$ on top of the GGE state $\ket{\rho}$~\cite{PhysRevLett.110.257203,DeNardis2018,CortesCubero2019}. We denote by $\Delta Q_e^{(\alpha)} = \bra{\rho; e} \bm{Q}^{(\alpha)} \ket{\rho; e} - 
\bra{\rho} \bm{Q}^{(\alpha)} \ket{\rho} $ the extra charge due to the excitation $e$ on top of $\ket{\rho}$~\cite{PhysRevLett.113.187203}. Similarly, $\Delta P_e$ is the momentum of the excitation $e$, while the matrix element $\bra{\rho} \bm{O} \ket{\rho; e}$ is a generalized form factor on top of the state $\ket{\rho}$~\cite{DeNardis2016,CortesCubero2019}. We also introduced the Fourier transform of the noise correlation $ \hat F(k) = \sum_j F(j) e^{- \imath k j}$.
We are now interested in the limit of smooth noise with finite correlation length. We thus parametrise $F(j) = \ell f(j/\ell)$, where $f(x)$ is an even and smooth function decaying to zero for $x\gg1$.
Expanding $\hat F(k)$ for $\ell\gg 1$, we have
\begin{equation}
\label{eq:expFT}
 \frac{\hat F(k)}{2\pi} = \ell f(0) \delta(k) + \frac{\kappa_2}{\ell}  \delta''(k) + O(\ell^{-3}),
\end{equation}
where to simplify the notation we set $\kappa_2 = -f''(0)/2 > 0$. Once Eq. \eqref{eq:expFT} is injected in Eq. \eqref{eq:chargevar}, we observe that only excitations at small exchanged momentum $\Delta P_e$ are relevant. In this limit, the form factor is dominated by a single particle-hole excitation~\cite{DeNardis2016,DeNardis2019} and one can replace
\begin{equation}
\label{eq:ph}
 \sum_e \stackrel{\Delta P_e \to 0}{\longrightarrow} \int d\ph d\pp (1 - n(\ph)) n(\pp) + \ldots,
\end{equation}
where the integral runs over the \textit{dressed} momenta $p$ of the particle $\pp$ and the hole $\ph$, with $\Delta P_e = \pp - \ph$. The dressed momentum and the rapidities are related via $dp = 2\pi\rho_t(\lambda) d\lambda$, where $\rho_t(\lambda)$ is the total root density, which counts the number of available modes~\cite{ZAMOLODCHIKOV1990695}. For non-interacting systems, $\rho_t(\lambda)$ is a fixed function, but in the presence of interactions, it is state-dependent and is related to $\rho(\lambda)$ via integral equations~\cite{yang1969thermodynamics}. The filling function is expressed as $n(p(\lambda)) = \rho(\lambda)/\rho_t(\lambda)$ and it fully specifies a stationary state. The right-hand side of Eq. \eqref{eq:ph} ensures that the momentum $\ph$ ($\pp$) is unoccupied (occupied).
In particular, the leading order in Eq. \eqref{eq:expFT} gives a vanishing contribution as $\Delta Q_e^{(\alpha)} =  O(\Delta P_e)$.
The second term instead gives a finite result, which can be entirely expressed in terms of the single particle-hole form factor in the limit of vanishing momentum \cite{10.21468/SciPostPhys.5.5.054,Cubero2019}
$
 \lim_{\pp \to \ph} \bra{\rho} \bm{O} \ket{\rho; \{\pp, \ph\}} = V^{\bf{O}}(\ph)
$,
where $V^{\bf{O}}$ is related to the expectation value of the operator on a generic stationary state $2\pi V^{\bf{O}}(p)=\delta\langle \bm{O}\rangle/\delta n(p)$ (see SM \cite{Note1}). If the noise is coupled to a conserved charge, one has the simple result $V^{\bm{q}}(p)= q^{\rm dr}(p)$ \cite{DeNardis2018,DeNardis2019}.
In the rapidity space, the dressed single-particle eigenvalue $q^{\rm dr}(\lambda)$ is determined solving the integral equation $(1 + T n)q^{\rm dr} = q $ (where the scattering shift $T$ is seen here as a linear operator in the space of $\lambda$ and $[1]_{\lambda,\lambda'}=\delta(\lambda-\lambda')$ is the identity operator). For the sake of simplicity, we make a little abuse of notation using the same symbol $q^{\rm dr}$ to denote both the dependence on rapidities and momenta. 
We point out that the dressed momentum introduced above is conventionally defined as the integral of the dressed derivative of the bare momentum $p_\text{bare}$, since $2\pi\rho_t(\lambda)=(\partial_\lambda p_\text{bare})^\text{dr}$.

The terms of order $\ell^{-3}$ in Eq. \eqref{eq:expFT} generate more complicated excitations as such as two particle-hole terms in Eq. \eqref{eq:ph}.
Restricting ourselves to the first non-trivial term, we can perform the integration over $\pp$, and by employing the completeness of the set of charges $\bm{Q}^{(\alpha)}$, equation \eqref{eq:chargevar} can be recast into a diffusion equation for the root density $\rho(\lambda)$ or equivalently for the filling function $n(p)$ \cite{Note1} describing the state at any time $t$:
\begin{equation}\label{eq:finaldiffusion2}
 \partial_t  n_t(p) = \frac{\kappa_2 \gamma }{\ell}
 \ \partial_p \Big( (V^{\bf{O}}_t( p))^2 \partial_p  n_t(p) \Big)  + O(\frac{\gamma}{\ell^3}) +  O(\gamma^2).
\end{equation}
This final equation has the simple form of diffusion in the space of dressed momenta $p$ and is the main result of our work. The remaining details of the noise can be completely re-absorbed defining a rescaled time $\tau = \kappa_2 \gamma t/\ell$. 
In the case of a generic driving, Eq. \eqref{eq:finaldiffusion2} holds in the limit $\ell,\gamma^{-1} \gg 1$. However, in the case where  $\bm{O}$ is chosen as a conserved charge, it is expected to hold for arbitrary $\gamma$, provided $\ell$ is chosen large enough. Indeed, for $\ell\to\infty$, driving with a conserved charge leaves the system unscathed, thus the $O(\gamma^2)$ term is absent and all higher order ones.
On the contrary, in the generic case diffusive corrections of order $\gamma^2/\ell^0$ coming from Fermi golden rule type of scatterings \cite{Mallayya2019,friedman2019diffusive} are expected. In the following, we will focus on the most relevant case where the operator O is a conserved density.
Note that the diffusion constant $\propto [V^{\bf{O}}_t( p)]^2$ is time-dependent, since it depends on the state $n_t(p)$ itself: the resulting equation is highly non-linear. Additionally, the mapping from momentum to rapidity space (where the dressing is defined) also evolves in time~(see SM \cite{Note1} and Ref. \cite{thomas2013numerical} for details about the numerical solutions). 

\paragraph{\it The interacting Bose gas. ---}
\begin{figure}[t!]
\includegraphics[width=0.49\columnwidth]{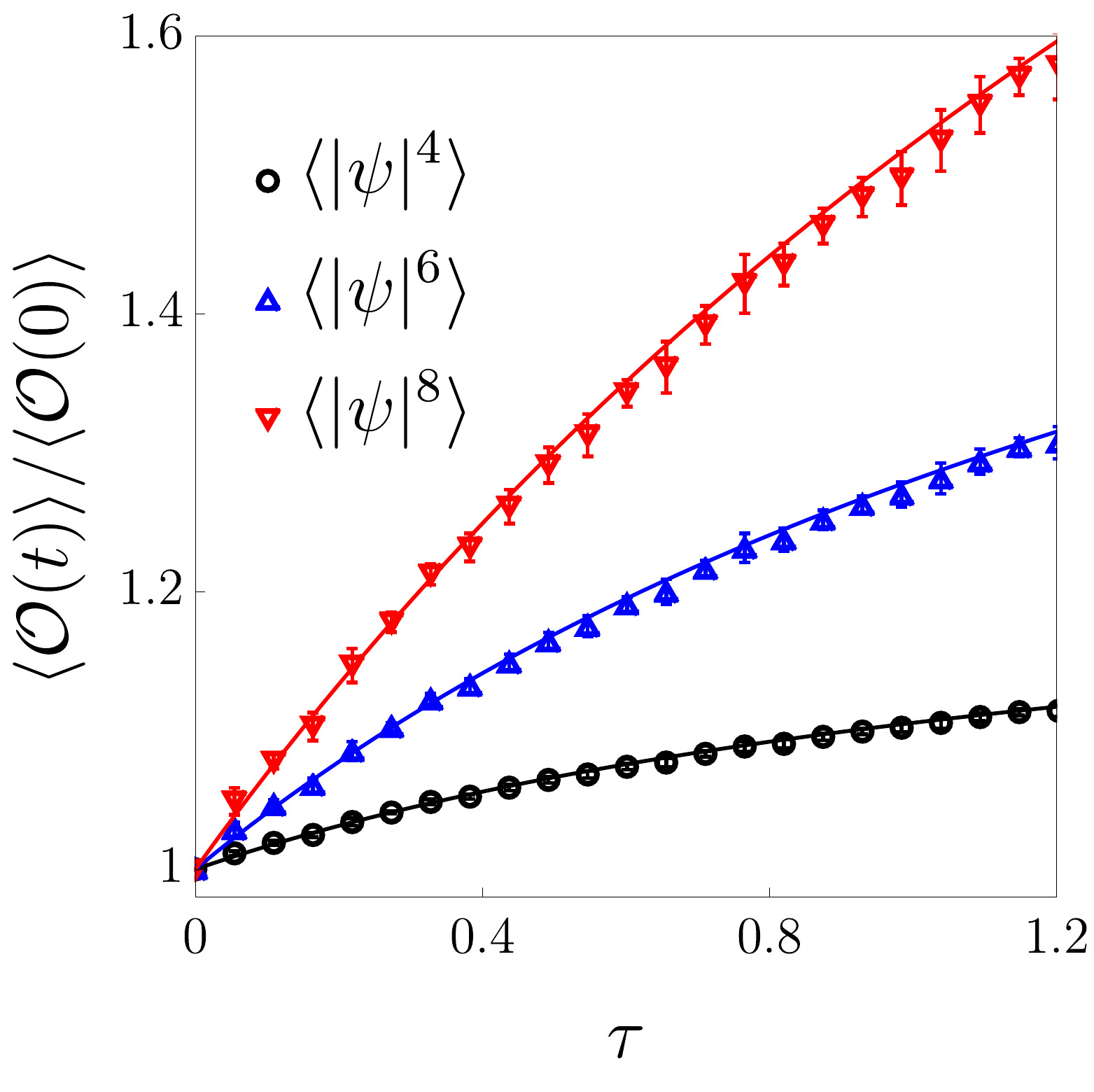}
\includegraphics[width=0.49\columnwidth]{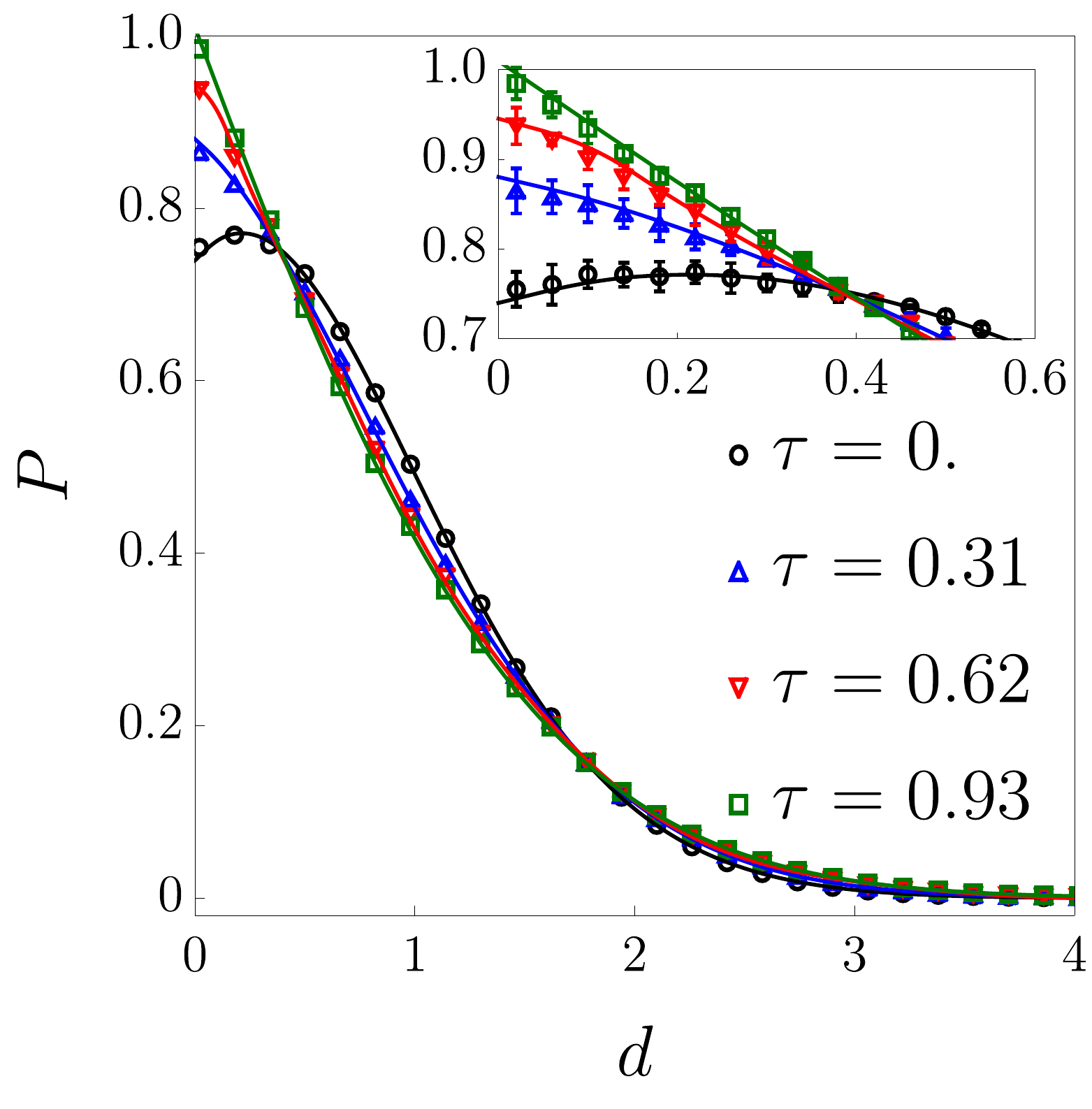}
\caption{\label{fig_NLS}Left: plot of the relative growth of the density moments $\mathcal{O}=\{|\psi|^4,|\psi|^6,|\psi|^8\}$ as a function of the rescaled time $\tau$ in the NLS.
Right: full counting statistics (FCS) of the particle density at different times (inset: zoom at small densities).
Solid lines: predictions from Eq. \eqref{eq:finaldiffusion2}. Symbols: ab-initio numerical simulations.
Above, the noise correlation is $F(x)=\ell \sqrt{\pi/2} e^{-x^2/(2\ell^2)}$ and the interaction $c=1$, the initial state is thermal with inverse temperature $\beta=1$ and chemical potential $\mu=2$ (resulting in density $\langle |\psi|^2\rangle=0.80$).
Agreement with the theoretical prediction is achieved with $\ell=4$ and $\gamma=0.1$.
}
\end{figure}
As a first application of our general findings, we revert to the 1d interacting Bose gas $\bm{H}_0=\int d x\, \partial_x\psi^\dagger\partial_x\psi+ c\psi^\dagger\psi^\dagger\psi\psi$, which is ubiquitous in describing the state-of-the-art cold atom experiments \cite{Bloch2005,RevModPhys.80.885,Kinoshita1125,PhysRevLett.95.190406,Kinoshita2006,PhysRevLett.105.230402,PhysRevLett.100.090402,PhysRevLett.122.090601}.
The model is integrable both in its classical \cite{Faddeev:1987ph} and quantum \cite{PhysRev.130.1605,PhysRev.130.1616} formulation. Continuous quantum models are notoriously hard to  simulate with tensor network techniques, hence its hydrodynamic description is a paramount achievement in experiments' simulations \cite{PhysRevLett.122.090601,mller2020introducing}.
Within the weakly-interacting regime and at finite temperature, the quantum system is well described by its classical limit \cite{doi:10.1080/09500340008232189,PhysRevLett.122.120401,PhysRevA.96.013623,PhysRevA.90.033611,PhysRevA.86.043626,PhysRevA.86.033626,PhysRevLett.122.120401}, i.e. the non-linear Schr\"odinger model (NLS), which is amenable to efficient ab-initio numerical simulations~\cite{10.1093/biomet/57.1.97,doi:10.1080/00031305.1995.10476177}. For this reason, hereafter we focus on the classical regime in the repulsive phase $c>0$ (see SM \cite{Note1} for details).
In Fig. \ref{fig_NLS}, we compare GHD predictions with numerical simulations, finding excellent agreement.
The system is initialized in a thermal state, which is then let to evolve with a noise coupled with the local density $\psi^\dagger(x)\psi(x)$.
Within the classical regime, the energy is not informative being UV divergent on thermal states \cite{10.21468/SciPostPhys.4.6.045}; therefore we consider the time evolution of the density moments $\langle|\psi(x)|^{2n}\rangle$ computed in Ref. \cite{vecchio2020exact} for arbitrary GGEs (see also Refs. \cite{PhysRevLett.120.190601,Bastianello_2018} for the quantum case).
Furthermore, we consider the evolution of the full counting statistics (FCS) of the density operator $P(d)=\langle \delta(\psi^\dagger(x)\psi(x)-d)\rangle$ \cite{vecchio2020exact}.
The details of the numerical simulations are left to SM \cite{Note1}.
With the chosen parameters, the initial thermal ensemble is strongly interacting, as confirmed by the anti-bouncing of the FCS \cite{vecchio2020exact}.
During the time evolution, the driving transfers energy from low momentum modes to more energetic ones, resulting in a progressive flattening of the filling with the consequence of diminishing the role of the interactions. Indeed, the FCS at late times approaches an exponential form $P(d) = d_0^{-1} e^{-d/d_0}$, with $d_0$ the average density, as expected in the non-interacting gaussian ensemble.

\paragraph{\it Interacting spin chains. ---} The XXZ spin chain is given by the Hamiltonian 
$
\bm{H}_0 = \sum_j \bm{h}_{j,j+1}
$
\begin{figure}[t!]
\includegraphics[width=0.99\columnwidth]{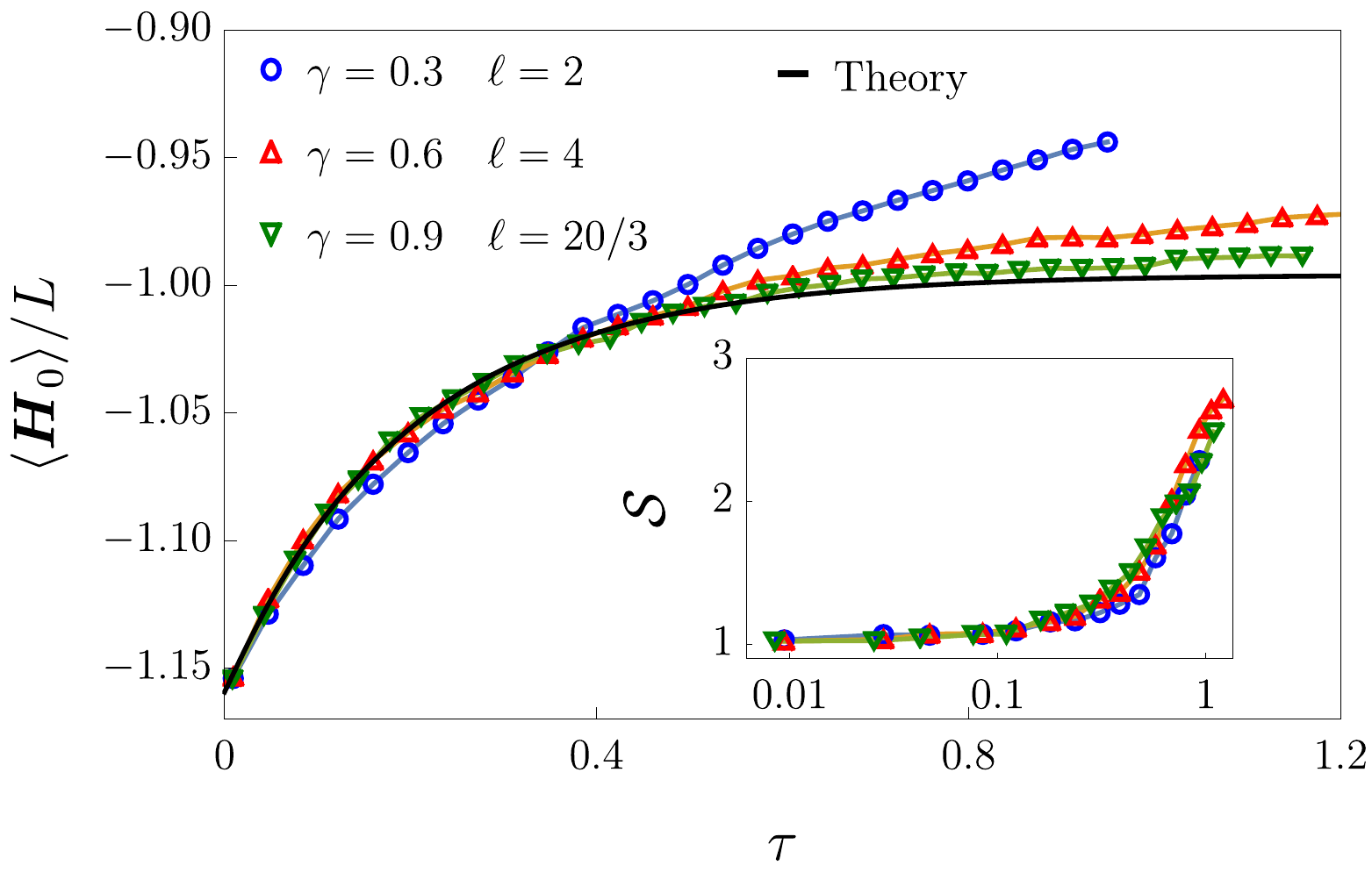}
\caption{ \label{fig_XXZ}
Plot of the time evolution of the energy density of an XXZ chain obtained by the 2-sites TDVP algorithm \cite{Haegeman} and averaged over $20$ realisations with fixed maximal bond dimension $\chi=200$ and $L=300$. The chain is initially prepared in the gapless groundstate at  $\Delta = \cosh (3/2)$ and $B=1.75946$, such that $\langle \bm{S}^z_j \rangle = 1/10$. The noise is coupled with the energy density $\bm{q}_j =\bm{h}_{j,j+1}$. Inset: log-plot of the entanglement entropy of the bi-partite chain. 
}
\end{figure}
where $\bm{h}_{j,j+1}= \bm{S}^x_j\bm{S}^x_{j+1} + \bm{S}^x_j \bm{S}^x_{j+1}+ \Delta (\bm{S}^z_j\bm{S}^z_{j+1}-1/4)$
with $\bm{S}^{x,y,z}$ being spin$-1/2$ operators. We focus on the easy-axis regime $\Delta>1$, where the  quasiparticle are labelled by an extra integer index $s =1,\ldots,\infty$ representing their spin quantum number. The ground state is formed by a Fermi sea of quasiparticles with spin$-1/2$, namely with $s=1$, while thermal states contain any spin $s$. Here, we restrict ourselves to the dynamics close to the ground state, so that only quasiparticles with $s=1$ are relevant. Since no new quasiparticles are generated by the dynamics of Eq. \eqref{eq:finaldiffusion2},
a gapped groundstate remains exactly unperturbed by our dynamics in the limit $\ell \to \infty$.
However, by adding an external magnetic feld  $\bm{H}_0 + B \sum_j \bm{S}^z_j $, we can choose $\Delta$ and $B$ such that the groundstate is gapless and the dynamics is non-trivial~\cite{takahashi2005thermodynamics, Bastianello_2019}. 
As an example, we drive the system with the local energy density $\bm{q}_j =\bm{h}_{j,j+1}$, which describes the effect of a phononic bath \cite{Lange2017,Lenari2018}.
We plot the evolution of the total energy of the system in Fig. \ref{fig_XXZ} and we notice that agreement with our theoretical prediction does indeed improve for large $\ell$, no matter the value of $\gamma$, as it is expected driving with a conserved charge. We observe that equation Eq. \eqref{eq:finaldiffusion2} predicts that energy increases up to a plateau on a pre-thermal stationary state, different from an infinite temperature state. The latter indeed cannot be reached by the dynamics given by Eq. \eqref{eq:finaldiffusion2}, as it requires creating quasiparticle with higher spin $s$, which are not contained in the initial state. Clearly, the corrections in $\ell^{-3}$ will include terms leading to quasiparticle production that will lead the system to thermalize. However in time scales of order $\ell/\gamma$ we observe perfect agreement of the numerical simulation with the evolution \eqref{eq:finaldiffusion2}, proving that it correctly describes the dynamics of the system at such time scales.

It is also interesting to consider the evolution \eqref{eq:finaldiffusion2} at short times. Starting from the ground state, this implies an initial linear growth in time for the charges, in particular, for the energy we have 
\begin{equation}\label{eq_short}
\frac{\langle \bm{H}_0 \rangle}{L} = e_{\rm GS} +  
{(V^{\bm O}(p_F))^2}{} v_F\  \tau/\pi + O(\tau^2)
\end{equation}
where $e_{\rm GS} $ is the ground state energy, $v_F$ is the Fermi velocity of the system and $p_F$ is the Fermi momentum.
In the case of the driving coupled to the local spin $\bm{q}_j  = \bm{S}^z$, we have $(V^{\bm O}(p_F))^2 = K$ the Luttinger parameter of the ground state, which recovers the prediction from bosonization~\cite{haldane1981demonstration,haldane1981luttinger,haldane1981effective,cazalilla2011one} (see SM \cite{Note1} for details).

\paragraph{\it Single noise realization. ---} 
The dynamics given by Eq.~\eqref{eq:finaldiffusion2} describes the average over several realizations of the noise in the evolution given by Hamiltonian~\eqref{eq:defModel} \cite{PhysRevLett.122.130605,PhysRevE.96.052118}. However, in each single realization, the evolution is pure and unitary, and the noise term plays the role of a random force. One can wonder whether Eq. \eqref{eq:finaldiffusion2}
can be recovered averaging the evolution induced by a stochastic external force.
In the case where the driving is coupled to an external charge, the hydrodynamic equations at first order in the external perturbation are known \cite{doyon2017note}. However, they are applicable in a regime of weak space-time dependence where the hydrodynamic picture applies. On the contrary, in our case the noise wildly fluctuates in time being it $\delta$-correlated. Nevertheless, since the effect of the noise is weak in our regime $\ell \gg 1$, we can assume that in each noise realization the system remains locally close to a quasi-equilibrium state. Then, the stochastic evolution of the stochastic filling function $\nn_{x,t} \equiv \nn_{x,t}(\lambda)$ reads~\cite{doyon2017note} 
\begin{equation}\label{eq:evolutionForce}
\partial_t \nn_{x,t} + v^{\rm eff}_{x,t} \partial_x \nn_{x,t}   -  (\partial_x U_{x,t}) \left( \frac{q^{\rm dr}_{x,t}} {p'_{x,t} }\right)\partial_\lambda \nn_{x,t} =  0 ,
\end{equation}
where $U_{x,t} = \sqrt{\gamma} \eta_{x}(t)$. The effective velocity of the quasiparticle is given by their \textit{dressed} energy $\varepsilon$, $v^{\rm eff} [\varepsilon]= \partial \varepsilon/\partial p = \partial_\lambda\varepsilon(\lambda)/\partial_\lambda p(\lambda)$, is also modified by the external force via $ \partial_\lambda \varepsilon \to  \partial_\lambda  \varepsilon + U_{x,t} (\partial_\lambda q)^{\rm dr}$. Note that the noise in Eq. \eqref{eq:evolutionForce} is meant in the Stratonovich convention which makes  the noise average non trivial. Nevertheless, after converting to the Ito formulation, one obtains a translational invariant filling $n_t(\lambda) \equiv \overline{\nn_{x,t}(\lambda)}$ whose time evolution in the space of momenta matches with Eq.~\eqref{eq:finaldiffusion2}.  

It is interesting to observe that Eq.~\eqref{eq:evolutionForce} does not lead to any entropy production~\cite{10.21468/SciPostPhys.6.6.070}. Starting from a Fermi sea distribution for $\nn_{x,t=0}(\lambda)$, the evolution \eqref{eq:evolutionForce} for a single realization of the noise can be seen as a local (random) boost of the Fermi points so that the state remains a zero-entropy one at all times. It is natural to expect this to result in the suppression of the entanglement entropy production. This is indeed what we observed in the tDMRG simulation at short times, see Fig. \ref{fig_XXZ}, where the growth of entanglement entropy is indeed curbed. However, at times of order $t \sim \kappa_2 \gamma/\ell $, the entanglement entropy of each realization suddenly starts growing linearly with time, signalling that its dynamics is given by terms that go beyond Eq. \eqref{eq:evolutionForce}, as for example diffusive terms of order $O(\partial_x^2\nn_{x,t})$ \cite{DeNardis2019}. However, Eq.~\eqref{eq:finaldiffusion2} still provides a good description of the averaged evolution of local operators. A similar entropy increasing at intermediate time scales was observed in the studies of classical hard rods in external potentials \cite{10.21468/SciPostPhys.6.6.070,PhysRevLett.120.164101}, where it was shown that the leading order hydrodynamic equation \eqref{eq:evolutionForce} is a good description also at intermediate times provided the space $\theta$ and $x$ are properly coarse-grained. 
\paragraph{\it Discussion and conclusion. ---} We presented an exact hydrodynamic description of the out-of-equilibrium dynamics of integrable systems in the presence of dephasing noise. 
A natural extension of our study would be considering inhomogeneous setups where ballistic transport and dephasing-induced diffusion can coexist on different timescales~\cite{10.21468/SciPostPhys.7.2.024, PhysRevB.99.174203}. 
For future perspectives, it would be interesting to extend our treatment to include subleading terms, which result from multiple particle-hole excitations: these terms would be responsible for the production/annihilation of quasiparticles and could eventually lead to thermalization. Moreover, it would be interesting to analyze operators which do not conserve particles' number and for which, even at the leading order, the form factor cannot be expanded in terms of particle-hole excitations. This would allow the study of systems in the presence of dissipative processes, coupled to a finite-temperature bath~\cite{Molmer:93}, and more generally of out-of-equilibrium steady states resulting from driven-dissipative dynamics~\cite{Lange2017}. Additionally, an important application would be the study of three-body losses, which are one of the leading effects in cold-atom setups involving quantum gases~\cite{PhysRevLett.122.090601,PhysRevLett.120.190601}. 
Another exciting direction is the inclusion of a genuinely quantum noise term, as modelled by the coupling to an ensemble of bosonic/fermionic quantum oscillators~\cite{CALDEIRA1983587, DIOSI1993517}. 
Finally, it would be interesting to describe the entanglement growth we observe, which currently exiles our hydrodynamic description.
\paragraph{Acknowledgements. ---} The MPS-based TDVP simulations were performed using the ITensor Library  \cite{ITensor}. We thank A. Nahum and D. Bernard for useful discussions.
AB acknowledges support from the European Research Council (ERC) under ERC Advanced grant 743032 DYNAMINT. J.D.N. is supported by the Research
Foundation Flanders (FWO).

\bibliography{biblio}

\newpage

\onecolumngrid
\newpage 

\appendix
\setcounter{equation}{0}
\setcounter{figure}{0}
\renewcommand{\thetable}{S\arabic{table}}
\renewcommand{\theequation}{S\arabic{equation}}
\renewcommand{\thefigure}{S\arabic{figure}}

\begin{center}
{\Large Supplementary Material \\ 
\titleinfo
}
\end{center}
\section{Definition of Ito/Stratonovich integrals and conversion relations}
We report here the standard definitions of the Ito and Stratonovich stochastic integrals over a Wiener process $W_t$~\cite{gardiner1985handbook}. For the Ito convention, we set for any sufficiently smooth function $g(t)$
\begin{subequations}
 \label{itostrato}
\begin{align}
 \int_0^t g(t) dW &\equiv \lim_{\delta t \to 0} \sum_i g(t_{i-1}) (W_{t_i} - W_{t_{i-1}})\;, \quad \mbox{Ito} \\
 \int_0^t g(t) \circ dW &\equiv \lim_{\delta t \to 0} \sum_i g\Bigl(\frac{t_{i-1} + t_{i}}{2}\Bigr) (W_{t_i} - W_{t_{i-1}})  \;, \quad \mbox{Stratonovich}
\end{align}
\end{subequations}
where $W_{t}$ denotes a Wiener process, with
\begin{equation}
 \langle W_t \rangle = 0 \;, \langle (W_{t} - W_{t'})^2 \rangle = |t-t'|
\end{equation}
One should notice the difference between the two cases: in the second one, the function $g$ is computed at the middle point between the two different times $t$ and $t+ dt$. The main consequence is that it is not statistically independent from the increment $dW(t) \sim W_{t + dt} - W_{t}$. Therefore, the Ito convention is preferred if one is interested in taking the noise average
\begin{equation}
 \langle g(t) dW_t \rangle = \langle g(t) \rangle \langle dW_t \rangle = 0\, .
\end{equation}

The two conventions give rise to different equation of motions. However it is possible to convert one into the other. Consider two prototypical stochastic equations of the form in the Ito-Stratonovich conventions with a single variable $X$
\begin{subequations}
\label{sde}
\begin{align}
& dX_t = f(X_t) dt + g(X_t) dW_t \label{ito}\\
& dX_t = \tilde f(X_t) dt + \tilde g(X_t) \circ dW_t \label{strato}\, ,
\end{align}
\end{subequations}
which in the integrated form they become respectively
\begin{align}
& X(t) = X(0) + \int_0^t \, ds \; f(X(s))  + \int_0^t g(X(s)) dW\\
& X(t) = X(0) + \int_0^t \, ds \; \tilde f(X(s))  + \int_0^t \tilde g(X(s)) \circ dW\, .
\end{align}
One can show that Eqs.~\eqref{sde} are actually describing the same stochastic process if 
\begin{equation}
\label{eq:itotostrato}
 \tilde g(X) = g(X) \;, \qquad \tilde f(X) = f(X) - \frac{1}{2} g(X) g'(X)\, .
\end{equation}

\subsubsection{Multi-dimensional case}
This formula admits a simple generalizaton to the vector case. We replace Eq. \eqref{sde} with
\begin{subequations}
\label{sdevec}
\begin{align}
& dX_\mu(t) = f_\mu(X_t) dt + g_{\mu,a}(X_t) dW_a(t) \label{itovec}\\
& dX_\mu(t) = \tilde f_\mu(X_t) dt + \tilde g_{\mu,a}(X_t) \circ dW_a(t) \label{stratovec}
\end{align}
\end{subequations}
where $X_\mu$ is now a $N$-component vector indexed by $\mu=1,\ldots N$ and $dW_a$ are $M$ independent Wiener processes
\begin{equation}
 \langle dW_a(t) dW_{a'}(t) \rangle = \delta_{aa'} dt\, .
\end{equation}

The sum over repeated indexes is assumed. In this case, one has the relations
\begin{equation}
\label{itostratovec}
 \tilde g_{\mu, a}(X) = g_{\mu,a}(X) \;, \qquad \tilde f_\mu(X) = f_\mu(X) - \frac{1}{2} \partial_\nu g_{\mu,a}(X) g_{\nu,a}(X) \, .
\end{equation}

\section{Derivation of the Lindblad equation}
It is useful to rewrite the noise by introducing a set of indepedent processes. Given the noise correlation
\begin{equation}
\langle \eta_j(t) \eta_{j'}(t') \rangle = \gamma F(j-j') \delta(t-t')
\end{equation}
we define a linear combinaton of the noise which diagonalises the correlation, i.e. 
\begin{equation}
 \eta_j(t) dt = \sqrt{\gamma}\sum_{j'} f(j-j') dW_j(t)  \;, \quad \sum_k f(j-k) f(j'-k) = F(j-j')
\end{equation}
which implies
\begin{equation}
\langle dW_j(t) dW_{j'}(t) \rangle = \delta_{j,j'} dt  \, .
\end{equation}
The Schr\"odinger equation in Eq. \eqref{stochschr} can then be written in the form
Eq. \eqref{stratovec} as
\begin{equation}
\label{stochschrstrato}
d\ket{\psi} = -\imath \hat{H}_{0}  \ket{\psi} dt -\imath \sqrt{\gamma}\sum_j  \tilde q_j  \ket{\psi} \circ dW_j(t) \;, \quad \tilde q_j = \left[\sum_{j'} f(j-j') \hat\opp_{j'}\right]\, ,
\end{equation}
where $\tilde q_j$ is a smoothened version of the local density $\hat \opp_j$ via the convolution Kernel $f(j)$. We can now apply Eq.~\eqref{itostratovec} to obtain the Ito form of the Schr\"odinger equation
\begin{equation}
    d\ket{\psi} =
    -\imath H_0 \ket{\Psi} dt -\frac  {\gamma}  2\sum_j \tilde \opp_j^2 \ket{\Psi} dt - \imath \sqrt{\gamma}\sum_j \tilde\opp_j   \ket{\Psi} dW_j(t)\, .
\end{equation}
From this, applying the Ito's lemma, we can get the average evolution of the density matrix
\begin{equation}
\frac{d\dd}{dt} = -\imath [\hat H_0, \dd] + \gamma \sum_j \left(\tilde \opp_j \dd \tilde \opp_j - \frac12 \{ \tilde \opp_j^2, \dd \}\right)\, ,
\end{equation}
which coincides with Eq.~\eqref{eq:lindblad} in the main text.

\section{Derivation of the hydrodynamic equation}

Here we provide a detailed derivation of the hydrodynamic equation \eqref{eq:finaldiffusion2}. As we discussed in the main text, in the limit of weak noise $\gamma\ll 1$ and large correlation length $\ell \gg 1$, for any charge $\bm{Q}^{(\alpha)}$, Eq. \eqref{eq:chargevar} can be written as
\begin{equation}
\lim_{L \to \infty} \frac{\Tr[\bm{Q}^{(\alpha)} \dot{\bm{\dd}}]}{L} = 
\int \lambda q^{(\alpha)}\partial_t \rho(\lambda)=-\frac{\gamma \kappa_2}{\ell}  \int d\lambda \, (\partial_\lambda q^{\rm dr}\,^{(\alpha)}(\lambda)) [V^{\bm{O}}(p(\lambda)]^2
 \partial_p n(p(\lambda))\, .
\end{equation}

We now integrate by parts using the definition of dressing
\begin{multline}
-\int d\lambda \,\partial_\lambda q^{\text{dr}(\alpha)}(\lambda)[V^{\bm{O}}(p(\lambda)]^2 \partial_p n(p(\lambda))=-\int d\lambda\int d \lambda' \partial_{\lambda'}q^{(\alpha)}(\lambda')(1+nT)^{-1}_{\lambda',\lambda}[V^{\bm{O}}(p(\lambda)]^2 \partial_p n(p(\lambda))=\\ \int d\lambda\int d \lambda' q^{(\alpha)}(\lambda')\partial_{\lambda'}\Bigg[(1+nT)^{-1}_{\lambda',\lambda}[V^{\bm{O}}(p(\lambda)]^2 \partial_p n(p(\lambda))\Bigg]\, ,
\end{multline}
then we use the completeness of the charges $\bm{Q}^{(\alpha)}$ to extract from the infinitely-many integral equations a differential equation for the root density.
In the rapidity space we obtain
\be\label{Seq_ghd_rho}
\partial_t \rho(\lambda)=\frac{\gamma \kappa_2}{\ell}\partial_{\lambda} \Bigg[(1+nT)^{-1}[V^{\bm{O}}]^2 \partial_p n\Bigg]\, .
\ee

This equation can be greatly simplified if written in terms of the filling. Indeed, using $\rho(\lambda)=n(p(\lambda))\rho_t(\lambda)$, one can readily show
\be
\partial_t \rho(\lambda)=(1+n T)^{-1}(\rho_t\partial_t n)\, ,
\ee
which leads to the equation
\be\label{eq_S21}
\partial_t n(p(\lambda))=\frac{\kappa_2 \gamma}{\ell\rho_t(\lambda)}\Bigg[\partial_\lambda( [V^{\bm O}]^2\partial_p n)-\partial_\lambda n(p(\lambda))T^{\text{dr}} ([V^{\bm O}]^2\partial_p n))\Bigg]\, .
\ee
Above, we used the definition of the scattering kernel $T(1 + n T)^{-1} = (1+ T n)^{-1} T = T^{\rm dr}$. We finally pass to the momentum space: while performing this operation, one should not forget the time dependence of the mapping between the rapidity and momentum
\be\label{eq_S22}
\partial_t n(p(\lambda))=\partial_t n(p)+\partial_t p(\lambda) \partial_p n(p)\, .
\ee
We now use that the total root density is the dressed derivative of the momentum
\be
\rho_t(\lambda)= \frac{\partial_\lambda p_\text{bare}(\lambda)}{2\pi}-\int d \lambda' T(\lambda,\lambda')\rho(\lambda')\, ,
\ee
where $p_\text{bare}(\lambda)$ is the bare momentum, which is state independent. Taking the time derivative of the above and using \eqref{Seq_ghd_rho} we get
\be
\partial_t\rho_t(\lambda)=\frac{\gamma \kappa_2}{\ell} \int d \lambda' T(\lambda,\lambda')\partial_{\lambda'} \Bigg[(1+nT)^{-1}[V^{\bm{O}}]^2 \partial_p n\Bigg]=\frac{\gamma \kappa_2}{\ell}\partial_\lambda T^{\text{dr}}[V^{\bm{O}}]^2 \partial_p n\, .
\ee
In the last passage, we integrated by parts (assuming that the boundary terms can be neglected) and used the symmetry of the kernel. 
Using $dp=2\pi \rho_t d \lambda$ and integrating the above equation, we find
\be\label{eq_S25}
\partial_t p(\lambda)= \int_{\lambda_0}^\lambda d\lambda' 2\pi\partial_t \rho_t(\lambda') = \frac{\gamma \kappa_2}{\ell}2\pi T^{\text{dr}}[V^{\bm{O}}]^2 \partial_p n\Big|_{\lambda}-\frac{\gamma \kappa_2}{\ell}2\pi T^{\text{dr}}[V^{\bm{O}}]^2 \partial_p n\Big|_{\lambda_0}\, .
\ee
Above, we explicitly wrote both the integration boundaries. In several instances, the rapidity $\lambda_0$ can be chosen in such a way that $2\pi T^{\text{dr}}[V^{\bm{O}}]^2 \partial_p n\Big|_{\lambda_0}=0$. For example, this is true for parity invariant states $\lambda\to -\lambda$ if we set $\lambda_0$. Assuming this simplification and combining Eqs. \eqref{eq_S22} and \eqref{eq_S25} in Eq. \eqref{eq_S21}, we obtain Eq. \eqref{eq:finaldiffusion2} of the main text.
For the numerical solution of the hydrodynamic equation, we found that the most stable and efficient approach was to solve directly Eq. \eqref{eq_S21} directly in the rapidity space. As it is typical in diffusion-like equations (involving second derivatives), the Crank-Nicholson algorithm \cite{thomas2013numerical} provided the sufficient stability to evolve the equation at large times.

\section{Noise average over the force field }

Within this section, we work in the rapidity space and for the sake of simplicity we define the filling $n(\lambda) \equiv n(p(\lambda))$  rather than the one in terms of the dressed momenta $p$. 
We consider the equivalent formulation of the hydrodynamic equation \eqref{eq_S21}  (in the case where the driving is coupled to a conserved charge $\bm O=\bm q$)

\begin{equation}\label{eq:motionolambda}
\partial_t n_{t}  = \frac{\kappa_2 \gamma}{\ell}
  \left[ \frac{1}{{{p}'} }  { \partial_\lambda \left(\frac{n'   ({{q}}^{\rm dr} )^2}{{{p}'}} \right)-   n' T^{\rm dr} \frac{(q^{\rm dr})^2}{p' }  n' }{}  \right]_t,
\end{equation}
where $n' = \partial_\lambda n_t(\lambda)$ and where the right hand side is evaluated at time $t$. Notice that we used the general notation
\begin{equation}
g_1 T^{\rm dr} g_2= \int d\lambda' g_1(\lambda) T^{\rm dr}(\lambda,\lambda') g_2(\lambda').
\end{equation}
  
We now wish to show that the dynamics given by a GHD with a random force, namely :
\begin{equation}
d n_{x,t}(\lambda) +  v^{\rm eff}_{x,t}[\varepsilon+U q] \partial_x n_{x,t}(\lambda) dt = \sqrt{\gamma}  \frac{q^{\rm dr}_{x,t}(\lambda)}{p'_{x,t}(\lambda)} \partial_\lambda n_{x,t}(\lambda)  \partial_x \sum_{x'} F({x-x'})dW_{x'}(t),
\end{equation}
where $v^{\rm eff}[\varepsilon+U q] =   v^{\rm eff}_{x,t}[\varepsilon] + v^{\rm eff}_{x,t}[q]$ is given as the sum of two effective velocities, one given by the dispersion relation of the quasiparticles $\varepsilon(p)$ and one given by the correction to the energy given by the external potential
\begin{equation}
v^{\rm eff}_{x,t}[\varepsilon](\lambda) =  \frac{ \varepsilon' (\lambda)}{p'(\lambda)} \quad ,\quad v^{\rm eff}_{x,t}[q]  (\lambda)  =  \frac{\sum_{x'} F({x-x'})\eta(x',t) (q')^{\rm dr}(\lambda) }{p'(\lambda)}  .
\end{equation}

In order to average over the noise, we first need to apply Eq.~\eqref{itostratovec} to convert the noise in the Ito form, where the transformations are now defined as 
\begin{align}
& f_{x,\lambda}(n) = -  v^{\rm eff}_{x,t}[\varepsilon]^{\rm eff}(\lambda) \partial_x n_{x,t}(\lambda) \\
& g_{x,\lambda, x'}(n) = - \sqrt{\gamma } v^{\rm eff}_{x,t}[q]^{\rm eff}(\lambda)  F({x-x'}) \partial_x n_{x,t}(\lambda) 
+\sqrt{\gamma}  \frac{q^{\rm dr}_{x,t}(\lambda)}{p'_{x,t} (\lambda)} \partial_\lambda n_{x,t}(\lambda)  \partial_x F({x-x'})
\end{align}
Therefore, relation \eqref{itostratovec} reduces to computing the following, neglecting terms proportional to $\partial_x n$, which will average to zero, 
\begin{align}
& \int d\lambda' dx' dx'' \frac{\delta g_{x,\lambda, x'}}{\delta n_{x'',t}( \lambda')} g_{x'',\lambda', x'}= \int dx'\Big(  - \sqrt{\gamma} {v^{\rm eff}_{x,t}[q]^{\rm eff}}(\lambda) F({x-x'}) \partial_x g_{x, \lambda, x'}  \nonumber \\& +   \sqrt{\gamma}  n'_{x,t}(\lambda) \int d\lambda' \frac{\delta [q^{\rm dr}_{x,t}(\lambda)/p'_{x,t}(\lambda)]}{\delta n_{x,t}(\lambda')}  F'({x-x'}) g_{x,\lambda',x'} -  \sqrt{\gamma} \frac{q^{\rm dr}_{x,t}(\lambda)}{p'_{x,t}(\lambda)} F'({x-x'}) \partial_\lambda g_{x,\lambda,x'} \Big),
\end{align}
where we denotes with $F'$ and $n'$ the derivative respect too $x$ or $\lambda$, respectively and we used 
\begin{equation}
\frac{ \delta n_{x,t}(\lambda)}{ \delta n_{x',t}(\lambda')} = \delta(\lambda-\lambda') \delta(x-x').
\end{equation} 
The functional derivative can be taken using the definition of dressing $q^{\rm dr} = (1+  T n)^{-1} q$ and $p' = (1 + T n)^{-1} p'_{\rm bare}$. We therefore have the following relation
\begin{equation}
  \frac{\delta [q^{\rm dr}/p'^{\rm dr}](\lambda)}{\delta n(\lambda')}  =   \frac{1}{p' (\lambda)} T^{\rm dr}(\lambda,\lambda')q^{\rm dr}(\lambda') + \frac{q^{\rm dr}(\lambda)}{(p'(\lambda))^2}T^{\rm dr}(\lambda,\lambda') p'(\lambda') .
\end{equation}
Putting all terms together and finally averaging over the noise using 
\begin{equation}
 \langle dW_x(t) dW_{x'}(t) \rangle = \delta_{xx'} dt,
\end{equation}
and the property
\begin{equation}
\int dx F(x) F''(x) = - \int (F'(x))^2 = - 2 \kappa_2 \gamma /\ell,
\end{equation}
we can write the evolution of the averaged occupation function, as in Eq. \eqref{strato}, as  
\begin{align}\label{eq:interme}
&\partial_t n_t = \frac{  \kappa_2 \gamma}{\ell}  \Big[ \frac{(q')^{\rm dr}  q^{\rm dr}}{(p')^2}  n' + \frac{q^{\rm dr}}{p'} \partial_\lambda \left( \frac{q^{\rm dr}}{p'}  n' \right) -   n' T^{\rm dr}   \frac{  {(q^{\rm dr})^2 }{ }  }{p'}  n'   -  \frac{n'  \ q^{\rm dr}  }{( p')^2} { T^{\rm dr} {q^{\rm dr} }  n' }  . \Big]_t
\end{align}
 In order to compare with Eq. \eqref{eq:motionolambda} we first need then to express $(q')^{\rm dr} $ in terms of $(q^{\rm dr})'$. This can be done using 
\begin{equation}
(1 + T n ) q^{\rm dr} =   q.
\end{equation}
So that by deriving left hand side and right hand side we obtain
\begin{equation}
(q^{\rm dr})'  + T^{\rm dr} n' q^{\rm dr} = (q')^{\rm dr} ,
\end{equation}
where we used that the scattering kernel is only a a function of the difference of rapidities $T(\lambda,\lambda') = T(\lambda - \lambda')$.
Therefore the following relation is valid 
\begin{equation}
 n' \frac{(q'^{\rm dr} ) q^{\rm dr}}{(p')^2}   = n' \frac{(q^{\rm dr} )' q^{\rm dr}}{(p')^2}  + \frac{n'  q^{\rm dr} }{(p')^2 }  {T^{\rm dr} n' q^{\rm dr} }{} .
\end{equation}
The second term in the above equation simplifies the last term in Eq. \eqref{eq:interme} and we  therefore obtain our final equation \eqref{eq:motionolambda}.

\section{Numerical methods for the classical interacting Bose gas} 

Here we shortly outline the method used for the numerical simulation of the classical interacting Bose gas.
The continuum system is discretized on a finite lattice of $N$ sites and lattice spacing $a$, periodic boundary conditions are assumed.
Then, the algorithm consists in three steps.
\begin{enumerate}
\item \emph{Sampling the thermal distribution with a Metropolis-Hasting method \cite{10.1093/biomet/57.1.97}}

We consider $N$ complex variables $\{\psi_i\}_{i=1}^N$ which are the discretized field. Then, we aim to sample thermal distributions, i.e. the probability of a certain field configuration $p[\{\psi_i\}_{i=1}^N]$ obeys
\be\label{eq_pr_met}
p[\{\psi_i\}_{i=1}^M]=\frac{1}{\mathcal{Z}}e^{- \beta E[\{\psi_j\}_{j=1}^N]}\, , \hspace{2pc} E[\{\psi_j\}_{j=1}^N]= a\sum_{j=1}^N \left\{\frac{|\psi_{j+1}-\psi_j|^2}{a^2}+c|\psi_j|^4-\mu|\psi_j|^2\right\}
\ee
Above, periodic boundary conditions are assumed, $\beta$ is the inverse temperature and $\mu$ the chemical potential. The partition function $\mathcal{Z}$ is needed for a correct normalization, but its value is not important.
Then, the field configurations are updated according with an ergodic dynamics $\{\psi_i\}_{i=1}^N\to \{\psi_i'\}_{i=1}^N$, whose fundamental step is divided into three parts 
\begin{enumerate}
\item Choose at random a lattice site $j$.
\item Change the field in position $j$ as $\psi_j\to \psi_j +\delta \psi_j$, with $\delta \psi_j$ a random gaussian variable of zero mean and variance $\Omega=\langle |\delta\psi_j|^2\rangle$. The variance $\Omega$ must be chosen in such a way that the acceptance rate (see below) is roughly $0.5$.
\item Compute the change in the Metropolis energy $\delta E= E[\{\psi_j'\}_{j=1}^N]- E[\{\psi_j\}_{j=1}^N]$.
If $\delta E<0$ the new configuration is accepted, otherwise it is accepted with probability $e^{-\beta\delta E}$.
\end{enumerate}
The Metropolis evolution is an ergodic process with respect to the probability \eqref{eq_pr_met}, which can be sampled picking field configurations along the Metropolis evolution.
\item \emph{Time evolve a sampled configuration with the microscopic equation of motion for a given noise's realization}

The discrete equation of motion (in the Ito convention) to be solved is
\be
i\partial\psi_j(t)=-a^{-2}(\psi_{j+1}(t)+\psi_{j-1}(t)-2\psi_j(t))-c|\psi_j(t)|^2\psi_j(t)- \eta_j(t)\psi_j(t)
\ee
with $\eta_j(t)$ the driving. We now discretize the time on a finite grid with spacing $d t$ and define the fields $\psi_j(s d t)\to \tilde{\psi}_j(s)$ and $\eta_j(sd t)\to\tilde{\eta}_j(s)$. 
The discrete noise is a random real gaussian variable with zero mean and correlation
\be
\langle \tilde{\eta}_j(s) \tilde{\eta}_{j'}(s') \rangle=\frac{\gamma}{a d t}\delta_{j,j'}\delta_{s,s'}F(a(j-j'))\, .
\ee
This scaling guarantees the correct continuous limit.
In order to obtain a stable evolution, we trotterize the dynamic evolving separately first with  the interaction and noise and then with the kinetic term. 
The first part of the evolution states
\be
\tilde{\psi}_j(s+1/2)=e^{-i dt (c|\tilde{\psi}_j(s)|^2+\tilde{\eta}_j(s))}\tilde{\psi}_j(s)\, .
\ee
The kinetic part is instead solved in the Fourier space $\varphi_j(s)=\sum_{j'}e^{i j j' 2\pi/N} \tilde{\psi}_{j'}(s)$, where it acts diagonally
\be
\varphi_j(s+1)=\exp\left\{-i 2a^{-2}\left[1-\cos\left(2\pi jN^{-1}\right)\right]\right\}\varphi_j(s+1/2)\, .
\ee

The above steps are then repeated to evolve the field up to the desired time.
\item \emph{Average over the initial conditions and the noise realizations.}
 
We took advantage of the translation symmetry and perform spacial averaging as well.
\end{enumerate}

In our simulations we used $a=0.06$, $N=2^{10}$ and $dt=0.0125$: this choice guarantees us a good convergence within the desired precision.
The error bars are estimated running four independent and identical samplings and considering the variance. We collected roughly $3000$ samples in total.

\section{The semiclassical limit of the 1d Bose gas}

The thermodynamics of the quantum 1d interacting Bose gas in nowadays a textbook topic \cite{takahashi2005thermodynamics}.
Within the repulsive phase $c>0$, the model is described by a single species of excitation (while in the attractive regime $c<0$ there exist infinitely many bound states) and the quantum scattering shift $T_\text{q}(\lambda,\lambda')$ is
\be
T_\text{q}(\lambda,\lambda')=-\frac{1}{2\pi}\frac{2c}{(\lambda-\lambda')^2+c^2}\, ,
\ee
from which the hydodynamics description follows. The energy, momentum and number of particles eigenvalues are $e(\lambda)=\lambda^2$, $p(\lambda)=\lambda$ and $N(\lambda)=1$ respectively.
On the contrary, the classical model at finite energy density is far less studied, therefore we provide a short recap building on the findings of Ref. \cite{vecchio2020exact}. The description of the classical 1d interacting Bose gas can be accessed through a proper \emph{semiclassical limit} of the quantum model: the latter is achieved restoring $\hbar$ in the quantum Hamiltonian \cite{vecchio2020exact}
\be
\hat{\bm{H}}_0=\int d x\,\left[\hbar^2\partial_x\psi^\dagger\partial_x\psi+c \hbar^4\,\psi^\dagger\psi^\dagger\psi\psi\right]\, ,
\ee
then taking the limit $\hbar\to 0$: classical physics emerges from the quantum one in the limit of small interactions and high occupation numbers, as it is expected.
Such a limit results in the classical scattering shift
\be
T_\text{cl}(\lambda,\lambda')=\lim_{\xi\to 0^+}-\frac{1}{2\pi}\frac{2c}{(\lambda-\lambda')^2+\xi}\, .
\ee
The kernel is formally singular and in integral expressions the limit $\xi\to 0^+$ must be taken only after the integration. For any fast decay test function $\tau(\lambda)$ this amounts to the regularization
\be
\int d \lambda' \, T_{\text{cl}}(\lambda,\lambda')\tau(\lambda')=\fint \frac{d\lambda'}{2\pi} \frac{2c}{\lambda-\lambda'}\partial_{\lambda'}\tau(\lambda')\, ,
\ee
where on the r.h.s. the singular integral is regularized with the principal value prescription. Albeit there is not a formulation of the form factors within the classical context, the classical hydrodynamics can be obtained through the semiclassical limit of the quantum one, using the correspondence $\lim_{\hbar\to 0} \hbar^2 n_\text{q}(\lambda) \tau^{\text{dr}(q)}(\lambda)=n_\text{cl}(\lambda) \tau^{\text{dr(cl)}}$ for an arbitrary test function $\tau(\lambda)$, where with $\tau^{\text{dr}(q)}$ and $\tau^{\text{dr}(cl)}$ we label the dressing performed with the quantum and classical kernel respectively.
The result of this limit is exactly Eq. \eqref{eq:finaldiffusion2}, where the dressing is performed with the classical kernel. The energy, momentum and particles eigenvalues remain $e(\lambda)=\lambda^2$, $p(\lambda)=\lambda$ and $N(\lambda)=1$ respectively.
In the main text, we chose as the initial state thermal ensembles, whose filling is determined by the following non-linear integral equation
\be
\varepsilon(\lambda)= \beta\big[e(\lambda)-\mu\big]-\int \frac{d \lambda'}{2\pi}\frac{2c}{\lambda-\lambda'}\partial_{\lambda'}\log\left( \varepsilon(\lambda')\right)
\ee
with $n_\text{cl}=1/\varepsilon(\lambda)$. Notice that with respect to the quantum case, the parametrization of the filling in terms of the pseudoenergy $\varepsilon(\lambda)$ is different: within the quantum case, the excitations follow a Fermi-Dirac distribution, while in the classical case one finds a Rayleigh-Jeans law. This is ultimately responsible for the UV divergence of the energy expectation value on thermal states, analogously to the famous UV-catastrophe in the black-body radiation.
As observables, we focused on the momenta of the density $|\psi(x)|^{2n}$ and on the FCS of the density operator: these can be determined on arbitrary GGEs (hence at any time of the GHD solution) solving proper integral equations. The general expressions can be found in Ref. \cite{vecchio2020exact}.

\section{The bosonization approach}

The Luttinger field theory \cite{haldane1981demonstration,haldane1981luttinger,haldane1981effective,
Cazalilla_2004} is ubiquitous in describing the ground state and low energy excitations of 1d systems with a $U(1)$ conserved quantity, regardless the integrability of the model. This charge could be, for instance, the number of particles in the interacting Bose gas or the $z-$magnetization in the XXZ spin chain.
Therefore, we can access the short-time behavior within bosonization, which must be in agreement with the short-time expansion of our hydrodynamic evolution.
Fluctuations of the conserved charge $\delta \bm{q}(x)$ are described in terms of a phase field $\hat{\phi}(x)$ with the correspondence $\delta \bm{q}(x)=-\frac{1}{\pi} \partial_x\hat{\phi}$: we focus only on a dephasing associated with the $U(1)$ charge, therefore the dynamics is governed by the following stochastic Hamiltonian
\be
\bm{H}_0=\frac{1}{2\pi}\int d x\,  v\Big[K \pi^2 \hat{\Pi}^2+\frac{1}{K}(\partial_x\hat{\phi})^2\Big]-\int d x\,  \eta_x(x) \frac{\partial_x\hat{\phi}}{\pi}+\text{const.}\, .
\ee
Above, $\hat{\Pi}$ is the momentum conjugated to the phase field $[\hat{\phi}(x),\hat{\Pi}(y)]=i\delta(x-y)$ and the noise $\eta_x(t)$ is, as usual, gaussian and $\delta-$correlated $\langle \eta_x(t)\eta_{x'}(t')\rangle=\gamma\delta(t-t')F(x-x')$. The parameters $v$ and $K$ are the sound velocity and the Luttinger parameter respectively: within generic models these can be numerically or experimentally extracted from the correlation functions, but in the case of integrable models they can be analytically determined \cite{Cazalilla_2004}.
In the absence of noise, the Luttinger Hamiltonian is diagonalized in terms of bosonic modes $[\hat{a}(k),\hat{a}^\dagger(q)]=2\pi \delta(k-q)$
\be
\hat{\phi}(t,x)=\int \frac{d k}{2\pi}e^{ik x}\sqrt{\frac{\pi K}{2 |k|}} \big[e^{-i v|k| t}\hat{a}(k)+e^{i v|k| t}\hat{a}^\dagger(-k)\big]\,, \,\,\,
\hat{\Pi}(t,x)=\int \frac{d k}{i2\pi}e^{ik x}\sqrt{\frac{|k|}{2\pi K}} \big[e^{-i v|k| t}\hat{a}(k)-e^{-i v|k| t}\hat{a}^\dagger(-k)\big]\, .
\ee
and the Hamiltonian is diagonalized as $\hat{\bm H}_0|_{\eta=0}=\int dk v|k| \hat{a}^\dagger(k)\hat{a}(k)$. Therefore, the ground state is the vacuum $\hat{a}(k)|0\rangle=0$.
In the presence of the noise, the fields evolve according to the equation of motion
\be
\partial_t \hat{\Pi}=\frac{1}{\pi} \frac{v}{K} \partial_x^2\hat{\phi}-\frac{\sqrt{\gamma}}{\pi}\partial_x \eta_{x}(t)\,, \hspace{2pc} \partial_t\phi=\pi Kv \hat{\Pi}\, ,
\ee
which are easily solved, leading to the simple result $
\langle \hat{a}^\dagger(t,k)\hat{a}(t,q)\rangle=\delta(k-q) t\frac{K\gamma}{2\pi} |k| \hat{F}(k)
$: the average is taken both with respect to the quantum expectation value and to the statistical fluctuations of the noise.
The average mode population can be now fed in the Hamiltonian, resulting in the following heating rate
\be
\frac{\langle \bm{H}_0 \rangle}{L} =e_{\rm{GS}}+t\frac{\kappa_2\gamma}{\ell}\frac{K}{\pi}v\, ,
\ee
which is in perfect agreement with the short-time expansion of the hydrodynamic prediction \eqref{eq_short}.

\section{The particle-hole form factors at coincident rapidities}

We now show the identity 
\be\label{Seq_top}
\lim_{\pp \to \ph} \bra{\rho} \bm{O} \ket{\rho; \{\pp, \ph\}} = V^{\bf{O}}(\ph)=\frac{1}{2\pi} \frac{\delta\langle \bm{O}\rangle}{\delta n(p)}
\ee
for an arbitrary local operator $\bm O$.
Let us consider a generic GGE parametrized with a generating function $w(\lambda)$
\be\label{Seq_TBA}
\varepsilon(\lambda)=w(\lambda)+\int \frac{d \lambda'}{2\pi} T(\lambda,\lambda')\log(1+ e^{-\varepsilon(\lambda')})\, ,
\ee
where the effective energy $\varepsilon$ parametrizes the filling as $n(\lambda)=1/(1+e^{\varepsilon(\lambda)})$.
Then, in Ref. \cite{10.21468/SciPostPhys.5.5.054} it has been shown that $V^{\bm O}$ can be obtained taking the variation of $\langle \bm O\rangle$ with respect to the filling
\be\label{Seq_Vo}
-\delta\langle \bm{O}\rangle=\int d\lambda\,\rho(\lambda)(1-n(\lambda)) V^{\bm{O}}(\lambda) (\delta w)^\text{dr}(\lambda)\, .
\ee
Above, we are slightly abusing the notation using the same name for functions defined in the rapidity and momentum space.
Let us take the variation of Eq. \eqref{Seq_TBA}
\be
\delta\varepsilon(\lambda)=\delta w(\lambda)-\int \frac{d \lambda'}{2\pi} T(\lambda,\lambda')n(\lambda')\delta \varepsilon(\lambda')\, ,
\ee
where we recognize the definition of dressing, i.e. $\varepsilon(\lambda)=(\delta w)^\text{dr}(\lambda)$. On the other hand, from the very definition of the filling we have
$\delta  n(\lambda)=-n(\lambda)(1-n(\lambda))\delta \varepsilon(\lambda)$. Combining these last two identities in Eq. \eqref{Seq_Vo} we find
\be
\delta\langle \bm{O}\rangle=\int d\lambda\, \rho_t(\lambda) V^{\bm{O}}(\lambda) \delta n(\lambda)\, .
\ee
Finally, changing variable from the rapidity to the momentum space $2\pi \rho_t(\lambda) d\lambda=dp$, Eq. \eqref{Seq_top} immediately follows.

\end{document}